\documentclass[reqno]{amsart}

\usepackage{amssymb}

\newtheorem{theorem}{Theorem}[section]
\newtheorem{proposition}[theorem]{Proposition}
\newtheorem{lemma}[theorem]{Lemma}

\newtheorem{definition}[theorem]{Definition}
\newtheorem{remark}[theorem]{Remark}

\numberwithin{equation}{section}

\DeclareMathOperator{\tr}{tr}

\DeclareMathOperator{\dist}{dist}

\newcommand{\pr}{\prime}

\renewcommand\L{\mathrm{L}}
\newcommand\R{\mathbb R}

\newcommand\Z{\mathbb Z}

\newcommand\X{\mathbf{X}}
\newcommand\Y{\mathbf{Y}}
\newcommand\beps{\boldsymbol{\varepsilon}}

\newcommand\J{\mathbb J}

\newcommand{\cX}{\mathcal{X}}

\newcommand{\cP}{\mathcal{P}}
\newcommand{\cQ}{\mathcal{Q}}
\newcommand{\cC}{\mathcal{C}}
\newcommand{\cJ}{\mathcal{J}}

\newcommand\e{\mathrm{e}}

\newcommand\eps{\varepsilon}
\newcommand{\vrho}{\varrho}

\newcommand{\la}{\langle}
\newcommand{\ra}{\rangle}

\renewcommand\P{\mathbb P}

\newcommand{\abs}[1]{\left\lvert #1 \right\rvert}

\newcommand{\scal}[1]{\la #1 \ra}

\newcommand\beq{\begin{equation}}
\newcommand\eeq{\end{equation}}

\begin{document}


\title[Attractive Poisson random Schr\"odinger operators]
{Localization at low energies for attractive Poisson random Schr\"odinger operators}

\author{Fran\c cois Germinet}
\address{ Universit\'e de Cergy-Pontoise,
D\'epartement de Math\'ematiques,
Site de Saint-Martin,
2 avenue Adolphe Chauvin,
95302 Cergy-Pontoise cedex, France}
 \email{germinet@math.u-cergy.fr}

\author{Peter D. Hislop}
\address{ Department of Mathematics, University of Kentucky, 
Lexington, KY 40506-0027, USA}
 \email{hislop@ms.uky.edu}

\author{Abel Klein}
\address{University of California, Irvine,
Department of Mathematics,
Irvine, CA 92697-3875,  USA}
 \email{aklein@uci.edu}

\thanks{P.H.  was  supported in part  by NSF Grant
DMS-0503784.}
\thanks{A.K was  supported in part by NSF Grant DMS-0457474.}

\begin{abstract}
We prove exponential and dynamical
localization at low energies for the  Schr\"oding\-er operator with an attractive Poisson random potential  in any dimension.
We also conclude that the eigenvalues in that spectral region of localization
have finite multiplicity.
\end{abstract}

\dedicatory{Dedicated to Stanislas Molchanov on the occasion of his $\, 65^{\mathrm{th}}$ birthday.}

\maketitle

\section{Introduction and main results} 

The motion of an electron moving in an amorphous medium where identical impurities have been randomly scattered, each impurity creating a local attractive  potential, is   described by a Schr\"odinger equation with Hamiltonian
\begin{gather}\label{PoissonH}
H_X :=  -\Delta + V_{X} \quad \text{on} \quad \L^{2}(\R^{d}) , \\
\intertext{where the potential is given by}
  V_{X}(x):= - \sum_{\zeta \in X} u(x - \zeta), \label{PoissonV}
\end{gather}
with $X$ being the location of the impurities and $-u(x - \zeta)\le 0$  the attractive potential created by the impurity placed at $\zeta$.  Since the impurities are randomly distributed, it is natural to model the configurations of the impurities by a Poisson process on $\R^{d}$  \cite{LGP,PF}.

The \emph{attractive Poisson Hamiltonian} is  the random Schr\"odinger operator $H_{\X}$ 
in \eqref{PoissonH} with $\X$   a Poisson process on $\R^d$ with density  $\varrho >0$; 
 $ V_{\X}$ being then  an   \emph{attractive Poisson random potential}. 
 The attractive  Poisson Hamiltonian $H_\X$ is an $\R^d$-ergodic family of
random self-adjoint operators;
it follows from standard results (cf.\ \cite{KM,PF})
that there exists fixed subsets of $\R$ so that the spectrum
of $H_\X$,  as well as the pure point,
absolutely continuous, and singular continuous components,
are equal to these fixed sets with probability one.

Poisson Hamiltonians have been known to have  Lifshitz tails, a strong indication of localization, for  quite a long time  \cite{DV,CL,PF,Klop97,Sz,KP,St99}. In particular, the existence of Lifshitz tails for  attractive  Poisson Hamiltonians is proved in  \cite{KP}. But up to recently localization was known only in one dimension \cite{Stolz};  
the multi-dimensional case remaining an open question (cf. \cite{LMW}). 
 
We have recently proven localization at the bottom of the spectrum for  Schr\"odinger operators with positive Poisson random potentials in arbitrary dimension \cite{GHK,GHK2}. We  obtained both exponential (or Anderson) localization and dynamical localization, as well as finite multiplicity of eigenvalues.
In this article we extend these results to attractive Poisson Hamiltonians,  proving  localization at low energies.

Localization has been  known for Anderson-type Hamiltonians \cite{HM,CH,Klop95,KSS,Klop02,GKgafa,AENSS}. In random amorphous media,  localization  was known for some Gaussian random potentials \cite{FLM,U,LMW}.  
In all these case there is an ``a priori'' Wegner estimate in all scales (e.g.,  \cite{HM,CH,Klop95,CHM,Kir,FLM,CHN,CHKN,CHK}).

 Bourgain and Kenig's proved
 localization for the Bernoulli-Anderson Hamiltonian, an   Anderson-type Hamiltonian where the  coefficients of the single-site potentials are Bernoulli random variables \cite{BK}.  They established  a Wegner estimate  by a multiscale analysis using  ``free sites" and a new quantitative version of unique  continuation which gives  a lower bound on eigenfunctions. Since they obtained weak probability estimates and had discrete random variables, they also introduced a new method to prove Anderson localization from
estimates on the finite-volume resolvents given by a single-energy multiscale analysis.  The new
method does not use the perturbation of singular spectra method nor
Kotani's trick as in \cite{CH,SW}, which requires random variables with bounded densities.  It is also not an energy-interval multiscale analysis as in \cite{vDK,FMSS,Kle}, which requires better probability estimates.

 To prove localization for Poisson Hamiltonians \cite{GHK2},  we exploited 
 the  probabilistic properties of Poisson point processes to use the  new ideas introduced by  Bourgain and Kenig  \cite{B,BK}

 Here we study  attractive single-site potentials, which we write as  $-u$, where $u$ is a  nonnegative, nonzero  $\mathrm{L}^{\infty}$-function on $\R^{d}$ with compact support, with 
 \begin{equation} \label{u}
u_{-}\chi_{\Lambda_{\delta_{-}}(0)}\le u \le u_{+}\chi_{\Lambda_{\delta_{+}}(0)}\quad \text{for some constants $u_{\pm}, \delta_{\pm}\in ]0,\infty[ $}.
\end{equation}
($\Lambda_{L}(x)$ denotes  the 
box of side $L$ centered at $x \in \R^{d}$.) 
It follows that $H_{X}$ is essentially self-adjoint on $C_{c}^{\infty}(\R^{d})$ and  $\sigma(H_\X)=\R$  with probability one \cite{CL,PF}.  We show that the conclusions of \cite{GHK2} hold in an interval of negative energies of the form $]-\infty, E_0(\vrho)]$ for some $E_0(\vrho)<0$.
We obtain both exponential (or Anderson) localization and dynamical localization, as well as finite multiplicty of eigenvalues.

 For a given set $B$, we let $\chi_{B}$ be its characteristic function,   $\cP_{0}(B)$ the collection of its countable subsets, and  $\#B$  its cardinality.  Given $X \in  \cP_{0}(A)$ and $A\subset B$, we set $X_{A}:=X \cap A$ and 
$N_{X}(A):= \# X_{A} $.    We write  $\abs{A}$ for the Lebesgue measure of  a Borel set $A \subset \R^d$.  We let 
 $\Lambda_{L}(x):=x+ \left(-\frac L 2,\frac L 2\right)^{d}$  be  the 
box of side $L$ centered at $x \in \R^{d}$.  By $\Lambda$ we will always denote some box  $ \Lambda_L(x)$ , with $\Lambda_{L}$ denoting a box  of side $L$.
We set
 $\chi_{x}:= \chi_{\Lambda_{1}}(x)$,   the characteristic  function of the
 box of side $1$ centered at
 $x \in \mathbb{R}^d$. We write
$\langle x \rangle := \sqrt{1+|x|^2}$, 
$T(x) :=\scal{x}^\nu$ for some fixed $\nu>\frac d 2$. By $C_{a,b, \ldots}$, $c_{a,b, \ldots}$, $K_{a,b, \ldots}$, etc., 
 will always denote some finite constant depending only on 
$a,b, \ldots$.

 A Poisson process on a Borel set  $B \subset \R^d$ with density $\varrho >0$ is a map $\X$ from a probability space  $(\Omega,\P)$ to  $\cP_{0}(B)$,
  such that  for each Borel set $A\subset B$ with $\abs{A}<\infty$ the random variable  $N_{\X}(A)$     has  Poisson distribution with mean  $\varrho |A|$,  i.e., 
 \beq
 {\P}\{N_{\X}(A)=k\}=\tfrac {(\varrho |A|)^k} {k!} \mathrm{e}^{-\varrho |A|}\quad \text{for $k=0,1,2,\dots$},
 \eeq
  and the random variables  $\{N_{\X}(A_{j})\}_{j=1}^{n}$ are independent for disjoint Borel subsets $\{A_{j}\}_{j=1}^{n}$ (e.g., \cite{King,Reiss}).

\begin{theorem}\label{thmpoisson}  Let $H_{\X}$ be an attractive Poisson Hamiltonian on  $\L^{2}(\R^{d})$ with density $\vrho >0$.  Then
there exists an energy $E_{0}=E_0(\vrho)<0$ for which  the following holds  ${\P}$-a.e.: The operator $H_\X$ has pure point spectrum  in $]-\infty,E_0]$  with exponentially localized eigenfunctions, and, if    $\phi$ is an eigenfunction of $H_\X$ 
with eigenvalue $E \in]-\infty,E_0]$ we have, with $m_{E} := \tfrac 1 8 \sqrt{ \tfrac 1 2 E_{0}-E} \le  m_{E_{0}}:= \tfrac 18 \sqrt{-  \tfrac 1 2 E_{0}}$, that
\begin{equation}\label{expdecay}
 \|\chi_x \phi\| \le C_{\X,\phi} \, e^{-m_{E}|x|} \quad \text{for all $x \in \R^{d}$}.
\end{equation}
Moreover, there exist   $\tau>1$ and  $s\in]0,1[$ such that for  all eigenfunctions $\psi,\phi$ (possibly equal) with the same eigenvalue  $E \in]-\infty,E_0]$ 
we have 
\begin{equation}\label{SUDEC}
\| \chi_x\psi\| \, \|\chi_y \phi\| \le C_{\X} \|T^{-1}\psi\|\|T^{-1}\phi\| \, e^{\scal{y}^\tau} e^{-|x-y|^s} \quad \text{for all $x,y \in \Z^{d}$}.
\end{equation}
 In particular, the eigenvalues of $H_\X$ in $]-\infty,E_0]$ have finite multiplicity,  and  $H_\X$ exhibits dynamical localization in $]-\infty,E_0]$, that is, for any $p>0$  we have
\begin{equation}\label{dynloc}
\sup_t \| \scal{x}^p e^{-itH_\X} \chi_{]-\infty,E_0]}(H_\X) \chi_0 \|^2_2 < \infty.
 \end{equation}
\end{theorem}

The proof of Theorem~\ref{thmpoisson}  relies on   the construction introduced in \cite{GHK2} for Poisson Hamiltonians, based on   the new multiscale analysis of Bourgain \cite{B} and Bourgain-Kenig \cite{BK} for Bernoulli-Anderson Hamiltonians.
Exponential localization follows as in \cite{BK}.
The decay of  eigenfunction correlations given  in (\ref{SUDEC}) then follows from \cite{GKsudec2} as in \cite{GHK2}.  Dynamical localization and finite multiplicity of eigenvalues are consequences of   (\ref{SUDEC}).

The Bourgain-Kenig multiscale analysis requires some detailed knowledge about the location of the impurities, as well as information on ``free sites'', and relies on conditional probabilities. To deal with these issues
and also handle the  measurability questions that appear for the Poisson process, in \cite{GHK2} we  performed a finite volume reduction in each scale as part of the multiscale analysis.  

In this note we review the basic construction of \cite{GHK2}, and apply it to  attractive Poisson Hamiltonians.  But since these are unbounded from below, we need to modify   the finite volume reduction  and the ``a priori'' finite volume estimates.

\section{Attractive Poisson Hamiltonians} \label{sectpre}

The  Poisson process $\X$ on
  $ \R^d$ with density  $\varrho$ is  constructed from a marked Poisson process as follows: Let  $\Y$ be a Poisson process on
 $ \R^d$ with density  $2\varrho $, and to each $\zeta \in Y$
associate a Bernoulli random variable $\beps_{\zeta}$, either $0$ or $1$ with equal probability, with $\beps_{\Y}=\{\beps_{\zeta}\}_{\zeta \in \Y}$ independent random variables.  Then $(\Y,\beps_{\Y})$ is a Poisson process with density $2 \rho$ on the product space $ \R^d \times \{0,1\}$,
the \emph{marked Poisson process};  its    underlying probability space will still be denoted by $(\Omega,\P)$.   (We use the notation 
 $(Y,\eps_{Y}):=\{(\zeta,\eps_{\zeta}); \, \zeta \in Y\} \in \cP_{0}(\R^{d}\times \{0,1\})$.)
  Define   maps $\cX,\cX^{\pr}\colon \cP_{0}( \R^d \times \{0,1\})\to  \cP_{0}( \R^d)$ by
\begin{equation}\label{XY}
\cX (\tilde{Z}):= \{ \zeta\in \R^{d} ; \, (\zeta,1) \in \tilde{Z}\} , \quad  \cX^{\pr} (\tilde{Z}):= \{ \zeta\in \R^{d} ; \, (\zeta,0) \in \tilde{Z}\},
\end{equation}
for all $\tilde{Z} \in  \cP_{0}( \R^d \times \{0,1\})$.  
Then the maps $\X,
\X^{\pr}\colon \Omega \to \cP_{0}( \R^d)$,  given 
 by 
\begin{equation}\label{XYomega}
\X:=\cX(\Y,\beps_{\Y}), \quad \X^{\pr}:=\cX^{\pr}(\Y,\beps_{\Y}),
\end{equation}
i.e., $\X(\omega)=\cX(\Y(\omega),\beps_{\Y(\omega)}(\omega))$,
$\X^{\pr}(\omega)=\cX^{\pr}(\Y(\omega),\beps_{\Y(\omega)}(\omega))$, are     Poisson processes on $ \R^d$ with density $\vrho$ (cf. \cite[Section~5.2]{King}, \cite[Example~2.4.2]{Reiss}), and we have  
 \begin{equation}\label{NXNY}
N_{\X}(A) + N_{\X^{\pr}}(A)=  N_{\Y}(A) \quad \text{ for all Borel sets  $A \subset \R^{d}$}.
\end{equation}

If $\X$ is a    Poisson process on $\R^{d}$ with density $\vrho$, then  $\X_{A}$ is a Poisson process on $A$ with density $\vrho$  for each Borel set  $A \subset \R^{d}$,   with  $\{\X_{A_{j}}\}_{j=1}^{n}$ being  independent Poisson processes   for disjoint Borel subsets $\{A_{j}\}_{j=1}^{n}$. Similar considerations apply to $\X^{\pr}$ and to 
the marked  Poisson process   $(\Y,\beps_{\Y})$,  with $\X_{A},\X^{\pr}_{A},\Y_{A},\beps_{\Y_{A}}$ satisfying  \eqref{XYomega}.

From now on we fix a probability space
 $(\Omega,\P)$  on which the Poisson processes $ \X$ and $\X^{\pr}$, with density $\vrho$,
and $\Y$, with density $2\vrho$, are defined, as well as the Bernoulli random variables $\beps_{\Y}$, and we have \eqref{XYomega}.  All events will be defined with respect to this probability space. 
 $H_{\X}$ (and $H_{\Y}$)  will always denote an   attractive Poisson Hamiltonian on  $\L^{2}(\R^{d})$ with density $\vrho >0$ ($2\vrho$),    as  in \eqref{PoissonH}-\eqref{u}.

  We start by 
showing that the attractive  Poisson Hamiltonian is self-adjoint and we have trace estimates needed in the multiscale analysis.

\begin{proposition}\label{propsa} The attractive Poisson Hamiltonians
$H_{\X}$ and  $H_{\Y}$ are essentially self-adjoint on $\cC^\infty_c(\R^d)$ with probability one.  In addition, we have
 \beq \label{HK}
\tr \{ T^{-1} \e^{-t H_{\X}} T^{-1} \}< \infty \quad  \text{for all $t>0$ \ $\P$-a.e.} ,
\eeq
and
\beq \label{HK2}
 \tr ( T^{-1} \chi_{]-\infty,E]}(H_{\X}) T^{-1} )
 < \infty    \quad \text{for all $E \in \R$ \ $\P$-a.e.}
\eeq
\end{proposition}

Since the potential is attractive, it may create infinitely deep wells.  This is controlled
by the following estimate.

\begin{lemma}\label{lembb}
Given a box $\Lambda$ we set
 \beq \label{Zdelta}
\J_{\delta_+,\Lambda}:= \left(\tfrac12\delta_+\Z^{d}\right)\cap \Lambda.
\eeq
There exists $L^\ast=L^\ast(d,\vrho,\delta_+)$, such that for any $L\ge L^\ast$ we have
\beq \label{proba+}
\P\left\{N_{\Y} (\Lambda_{2\delta_+}(j)))\le \vrho\log L \quad \forall j\in\J_{\delta_+,\Lambda_L}\right\} \ge 1 - L^{- \frac 2 3 \log \log L}
\eeq
and
\beq\label{controlVL}
\P\{\|\chi_{\Lambda_L} V_{\Y}\|_\infty \le   u_+ \vrho\log L\} \ge 1 - L^{- \frac 2 3 \log \log L}.
\eeq
It follows that for $\P$-a.e.\ $\omega$ we have
\begin{equation}\label{controlV}
V_{\X(\omega)}(x)\ge V_{\Y(\omega)}(x)\ge -c_\omega \log (1+|x|) \quad \text{for all $x\in\R^d$},
\end{equation}
where  $c_\omega>0$ (depending also on $p,\delta_+,\vrho, u_+$). 
\end{lemma}

\begin{proof} We may assume $\vrho \log L \ge 1$.
Standard bounds on Poisson random variables (cf. \cite[Eq. (2.7)]{GHK2}) give
\beq 
\P\{N_{\Y} (\Lambda_{2\delta_+}(x))> \vrho\log L\} \le  \left(\tfrac{2\vrho(2\delta_+)^d}{\vrho\log L}\right)^{\vrho\log L} \le   L^{- \frac 3 4  \log \log L}
\eeq
for any $x\in\R^d$,  if $L\ge L^\ast_{1}(d,\vrho,\delta_+)$.  It follows that for
 $L\ge L^\ast(d,\vrho,\delta_+)$ we have
\beq 
\P\left\{ N_{\Y} (\Lambda_{2\delta_+}(j))\le \vrho\log L  \quad \forall j\in\J_{\delta_+,\Lambda_L}\right\} \ge 1- \left(\tfrac{2L}{\delta_+}\right)^{d} L^{- \frac 3 4 \log \log L} \ge 1 -  L^{- \frac2 3 \log \log L}.
\eeq
Now, for any $x\in\Lambda_L$ there exists $j\in\J_{\delta_+,\Lambda_L}$ s.t. $\Lambda_{\delta+}(x)\subset \Lambda_{2\delta+}(j)$. Hence, we also have
\beq \label{newe}
\P \{ N_{\Y} (\Lambda_{\delta_+}(x)\cap \Lambda_L) \le \vrho\log L  \quad \forall x\in\Lambda_L \} \ge1 -  L^{- \frac 2 3 \log \log L}.
\eeq
But if the event in \eqref{newe} occurs, it follows from \eqref{u} that
 \beq \label{lowerboundL}
|V_{\Y}(x)| \le   u_+ \vrho\log L \quad     \forall x\in\Lambda_L,
\eeq
and \eqref{controlVL}  follows, since
\beq
0 \ge V_{\X}(x) \ge V_{\Y}(x)  
\eeq
because of \eqref{XYomega}.
The Borel-Cantelli Lemma now gives \eqref{controlV}.
\end{proof}

In the one-dimensional case an estimate similar to \eqref{controlV} can be found in \cite{HW}.

\begin{proof}[Proof of Proposition~\ref{propsa}.]
In view of  \eqref{controlV}, it follows from the  Faris-Levine Theorem \cite[Theorem~X.38]{RS2} that  $H_{\X}$ and  $H_{\Y}$ are  essentially self-adjoint on $\cC^\infty_c(\R^d)$ with probability one.

The trace estimate \eqref{HK}  follows from Gaussian bounds on   heat kernels  \cite[Lemma~1.7]{BLM}, 
which hold  $\P$-a.e.  in view of \eqref{controlV}.  As a consequence, we have
\begin{align}
& \tr ( T^{-1} \chi_{]-\infty,E]}(H_{\X}) T^{-1} )
 = 
\tr ( T^{-1} \e^{-H_\X} \e^{2H_\X}\chi_{]-\infty,E]}(H_{\X}) \e^{-H_\X} T^{-1} ) \\
& \qquad \qquad \qquad \quad \le 
\e^{2E} \tr ( T^{-1} \e^{-2H_\X} T^{-1} ) < \infty    \quad \text{for all $E \in \R$ \ $\P$-a.e.}
\notag
\end{align}
\end{proof}

Given  two disjoint  configurations $X,Y \in \cP_{0}(\R^{d})$ and $t_{Y}=\left\{t_{\zeta} \right\}_{\zeta \in Y} \in [0,1]^{Y}$, we set 
 \begin{equation} \label{PoissonHY}
H_{X,(Y, t_{Y})}:=  -\Delta + V_{X,(Y, t_{Y})} , \ \text{where} \ V_{X,(Y, t_{Y})}(x) := V_{X}(x) + \sum_{\zeta \in Y} t_{\zeta}u(x - \zeta).
\end{equation}
In particular, given $\eps_{Y}\in \{0,1\}^{Y}$ we have, recalling \eqref{XY}, that
  \begin{equation}\label{PoissonHYeps}
H_{X,(Y, \eps_{Y})}= H_{ X \sqcup \cX(Y,\eps_{Y})}.
\end{equation}
We also write $H_{(Y, t_{Y})}:=H_{\emptyset,(Y, t_{Y})}$
and 
\begin{equation} \label{Homega}
H_{\omega}:= H_{\X(\omega)}= H_{(\Y(\omega),\beps_{\Y(\omega)}(\omega))}.
\end{equation}

\section{Finite volume}  \label{secfinvol}

The multiscale analysis requires finite volume operators,which  are defined as follows.  Given   a box  $\Lambda= \Lambda_{L}(x)$  in $\R^{d}$ and a configuration $X$, we set 
\begin{align}\label{finvolH}
H_{X,\Lambda} :=-{\Delta_{\Lambda}}+ V_{X,\Lambda} \quad \text{on}   \quad \L^{2}(\Lambda),
\end{align}
where $\Delta_{\Lambda}$ is the  the Laplacian on $\Lambda$ with Dirichlet boundary condition,  and  
\begin{equation}
 V_{X,\Lambda}:= \chi_{\Lambda} V_{X_{\Lambda}} \quad \text{with $V_{X_\Lambda} $ as in \eqref{PoissonV}}.  \label{finvolV}
\end{equation}
The finite volume resolvent  is 
 $R_{X,\Lambda} (z):=(H_{X,\Lambda} - z)^{-1}$.

We have 
$\Delta_{\Lambda}= \nabla_{\Lambda}\cdot\nabla_{\Lambda}$, where 
$\nabla_{\Lambda}$ is the gradient with Dirichlet boundary condition.
We sometimes identify $\L^{2}(\Lambda)$ with $\chi_{\Lambda }\L^{2}(\R^{d})$
and, when necessary,  will use subscripts $\Lambda$ and $\R^{d}$ to distinguish between the norms and inner products of $\L^{2}(\Lambda)$ and $\L^{2}(\R^{d})$.
Note that  we always have
\begin{equation}
 \chi_{\widehat{\Lambda}} V_{X,\Lambda}=  \chi_{\widehat{\Lambda}} V_{X,{\Lambda^{\prime}}}, 
\end{equation}
where
\beq    \widehat{\Lambda}=\widehat{\Lambda}_{L}(x):=\Lambda_{L-\delta_{+} }(x) \quad  \text{with $\delta_{+}$ as in \eqref{u}},  \label{lambhat}
\eeq
which suffices for the multiscale analysis.

The multiscale analysis estimates probabilities of desired properties of finite volume resolvents at  energies  $E\in \R$. ($L^{p\pm}$ means $L^{p\pm \delta}$ for some small $\delta >0$.    We will write  $\sqcup$ for disjoint unions: $C=A\sqcup B$ means  $C=A \cup B$ with  $A \cap B =\emptyset$.)

  \begin{definition} Consider an energy $E\in \R$ and a rate of decay $m>0$.  A   box  $\Lambda_{L}$ is said to be $(X,E,m)$-good if 
  \begin{align}\label{weg}
\| R_{X,\Lambda_{L}}(E) \|& \le \e^{L^{1-}}
\intertext{and} 
\| \chi_x R_{X,\Lambda_{L}}(E) \chi_y \|& \le \e^{-m |x-y|}, \;\;  \text{for $x,y \in \Lambda_{L}$ with $ |x-y|\ge \tfrac L{10}$}. \label{good}
\end{align}
 We say that the box $\Lambda_{L}$ is $(\omega,E,m)$-good if it is $(\X(\omega),E,m)$-good.  
 \end{definition}

Condition \eqref{good} is the standard notion of a regular box in the multiscale analysis \cite{FS,FMSS,vDK,GK1,Kle}.  Condition \eqref{weg} plays the role of a Wegner estimate. This control on the resolvent is just good enough so that it does not destroy the exponential decay obtained with \eqref{good}. Since Wegner is proved scale by scale (as in \cite{CKM,B,BK}), it is incorporated in the definition of the goodness of a given box.
    
But  \emph{goodness} of boxes does not suffice for the induction step in the multiscale analysis given in \cite{B,BK}, which also needs an adequate supply of \emph{free sites}  to obtain a Wegner estimate at each scale.   Given  two disjoint  configurations $X,Y \in \cP_{0}(\R^{d})$ and $t_{Y}=\left\{t_{\zeta} \right\}_{\zeta \in Y} \in [0,1]^{Y}$, we recall \eqref{PoissonHY} 
and define the corresponding finite volume operators $H_{X,(Y, t_{Y}),\Lambda}$ as in \eqref{finvolH}  and \eqref{finvolV} using $X_{\Lambda}$,  $Y_{\Lambda}$ and $t_{Y_{\Lambda}}$, i.e.,
\beq
H_{X,(Y, t_{Y}),\Lambda} := -\Delta_{\Lambda}+ V_{X,(Y, t_{Y}),\Lambda} \quad \text{with} \quad  V_{X,(Y, t_{Y}),\Lambda}:= \chi_{\Lambda} V_{X_{\Lambda},(Y_{\Lambda}, t_{Y_{\Lambda}})},
\eeq
 with 
 $R_{X,(Y, t_{Y}),\Lambda} (z)$ being the corresponding finite volume resolvent.

\begin{definition} Consider  two   configurations $X, Y \in \cP_{0}(\R^{d})$ and an energy $E$. A  box  $\Lambda_{L}$ is said to be $(X,Y,E,m)$-good if $X\cap Y=\emptyset$ and 
 we have   \eqref{weg} and  \eqref{good} 
with  $R_{X,(Y, t_{Y}),\Lambda_{L}} (E)$ for all $t_{Y} \in [0,1]^{Y}$.  In this case $Y$ consists of $(X,E)$-free sites for the box  $\Lambda_{L}$. \emph{(In particular, the box  $\Lambda_{L}$ is  $(X\sqcup \cX(Y,\eps_{Y}),E,m)$-good for all  $\eps_{Y} \in\{0,1\}^{Y}$.)}
\end{definition}

The multiscale analysis  requires some detailed knowledge about the location of the impurities, that is, about the Poisson process configuration, as well as information on ``free sites''. To deal with this and also handle the measurability questions that appear for the Poisson process, a finite volume reduction was performed in \cite{GHK2} as part of the multiscale analysis.  The key is  that  a Poisson point can be moved a little bit without spoiling the goodness of boxes \cite[Lemma~3.3]{GHK2}.
We now  recall  the construction of \cite{GHK2}, with some slight adaptations to our present setting.
  
  \begin{definition}  \label{deflat}   Let  $\eta_{L}:= \e^{-L^{10^{6}d}}$ for $L>0$.  Given a box $\Lambda=\Lambda_{L}(x)$,  set
  \begin{equation}
   \J_{\Lambda}:= \{ j \in x +\eta_{L}\Z^{d}; \ \Lambda_{\eta_{L}}(j) \subset \Lambda\}.
\end{equation}
 A configuration $X \in \cP_{0}(\R^{d})$ is said to be $\Lambda$-acceptable if (recall \eqref{Zdelta})
  \begin{gather}\label{totalN1}
 N_{X}(\Lambda_{2\delta_+}(j)) < \vrho \log L \quad \forall j\in\J_{\delta_+,\Lambda},\\
  N_{X}(\Lambda_{\eta_{L}}(j ))\le 1 \quad \text{for all} \quad j \in \J_{\Lambda}, \label{tinyN}
\\
     \intertext{and}
      \label{bryN}
   N_{X}(\Lambda\backslash \cup_{j \in  \J_{\Lambda} }\Lambda_{\eta_{L}(1-\eta_{L})}(j))=0;
 \end{gather}
it is $\Lambda$-acceptable$\,^{\prime}$ if it  satisfies \eqref{totalN1},\eqref{tinyN}, and the weaker
 \beq  \label{bryN2}
   N_{X}(\Lambda\backslash \cup_{j \in  \J_{\Lambda} }\Lambda_{\eta_{L}}(j))=0.
 \eeq
We set
\begin{align}
 \cQ_{\Lambda}^{(0)}:&=\{X \in \cP_{0}(\R^{d}); \, \text{$X$ is  $\Lambda$-acceptable}\},\\
  \cQ_{\Lambda}^{(0\prime)}:&=\{X \in \cP_{0}(\R^{d}); \, \text{$X$ is  $\Lambda$-acceptable$\,^{\prime}$}\},
\end{align}
 and consider the event (recall that $\Y$ is the Poisson process with density $2\vrho$)
 \beq \label{defOmega0}
\Omega_{\Lambda}^{(0)}:=\{\Y \in \cQ_{\Lambda}^{(0)}\} .
\eeq
\end{definition}

Note that  $ \Omega_{\Lambda}^{(0)} \subset \{\X\in \cQ_{\Lambda}^{(0)}\}$  and  $\cQ_{\Lambda}^{(0)} \subset \cQ_{\Lambda}^{(0\prime)}$. Condition \eqref{bryN} is put in to avoid ambiguities in the multiscale analysis.

\begin{remark}\label{rembconfset} Note that  \eqref{totalN1} is not the same as \cite[Eq. (3.19)]{GHK2}, and hence the above definitions of $\Lambda$-acceptable
and $\Lambda$-acceptable$\,^{\prime}$ configurations are slightly different from the ones given in \cite[Definition~3.4]{GHK2}. The reason for  \eqref{totalN1}  is to ensure the estimates \eqref{controlV2} and \eqref{lowerboundinbox} in the next lemma.
\end{remark}

\begin{lemma}\label{lemlat5}   Let $\Lambda= \Lambda_{L}$.   Then for all
 $\Lambda$-acceptable$\,^{\prime}$ configurations $Y$ we have
\begin{gather}
N_{Y}(\Lambda)\le \vrho (2\delta_+)^{-d}L^{d }\log L,
 \label{totalN} \\
\label{controlV2}
V_{Y}(x)\ge -u_{+}\vrho \log L \quad \text{for all  $x\in\Lambda$},\\
\intertext{and}
\label{lowerboundinbox}
 H_{(Y,t_{Y}),\Lambda} \ge  -u_{+}\vrho \log L \quad\text{for all  $t_{Y} \in [0,1]^{Y}$}.
  \end{gather}
\end{lemma}

\begin{proof}The estimate \eqref{totalN} is an immediate consequence of \eqref{totalN1}.  It also follows from \eqref{totalN1}, by the same argument used for \eqref{lowerboundL}, that  for all  $\Lambda$-acceptable$\,^{\prime}$ configurations $Y$ we have the lower bound \eqref{controlV2},
from which we get \eqref{lowerboundinbox} since $ H_{(Y,t_{Y}),\Lambda}\ge  H_{Y,\Lambda}\ge V_{Y,\Lambda}$ .
\end{proof}

\begin{lemma}\label{lemlat}   
 There exists a scale $\overline{L}=\overline{L}(d,\vrho,\delta_{+}) < \infty$, such that if $L\ge \overline{L} $ 
we have
 \begin{equation}
\P \{\Omega_{\Lambda_{L}}^{(0)}\} \ge 1 - L^{- \frac 1 2 \log \log L} .
 \label{Omega0}
\end{equation}
\end{lemma}

\begin{proof}   Recalling \eqref{proba+}, we have
 \begin{equation}
\P \{\Omega_{\Lambda_{L}}^{(0)}\} \ge 1 - L^{- \frac 2 3 \log \log L} - 4d \vrho  (L^{d-1}+L^{d})\eta_{L} -  2 \vrho^{2} L^{d} \eta_{L}^{d},
\end{equation}
and hence \eqref{Omega0} follows for large $L$.
\end{proof}

Lemma~\ref{lemlat} tells us that  inside the box $\Lambda$, outside an event of  negligible probability in the multiscale analysis, we only need to consider $\Lambda$-acceptable configurations of the Poisson process $Y$. 

Fix a box  $\Lambda=\Lambda_{L}(x)$, then
\beq
X\overset{\Lambda}{\sim} Y \Longleftrightarrow  N_{X}(\Lambda_{\eta_{L}}(j ))=N_{Y}(\Lambda_{\eta_{L}}(j )) \quad \text{for all} \quad j \in \J_{\Lambda}\label{equivrel}
\eeq
introduces an equivalence relation  in both $\cQ_{\Lambda}^{(0\prime)} $ and  $\cQ_{\Lambda}^{(0)} $;  the equivalence class of $X$ in $\cQ_{\Lambda}^{(0\prime)} $ will be denoted by $[X]_{\Lambda}^{\prime}$. If $X \in \cQ_{\Lambda}^{(0)}$, then  $[X]_{\Lambda}=[X]_{\Lambda}^{\prime}\cap \cQ_{\Lambda}^{(0)} $ is its  equivalence
class in $\cQ_{\Lambda}^{(0)}$. 
 Note that $[X]_{\Lambda}^{\prime}=[X_{\Lambda}]_{\Lambda}^{\prime}$. We also
write\beq  \label{[A]}
 [A]_{\Lambda}: = \bigcup_{X \in A} [X]_{\Lambda} \quad \text{for subsets}\quad  A \subset \cQ_{\Lambda}^{(0)}.
\eeq

The following lemma \cite[Lemma~3.6]{GHK2}  tells us that ``goodness'' of boxes is a property of equivalence classes of acceptable$\,^{\prime}$ configurations: changing configurations inside an equivalence class takes
good boxes  into just-as-good (jgood) boxes. Proceeding as in the lemma, we find that changing configurations inside an equivalence class takes jgood boxes  into what we may call just-as-just-as-good (jjgood) boxes, and so on.  Since we will only carry this procedure a bounded number of times, the bound independent of the scale, we will simply call them all  jgood boxes.

\begin{lemma}[\cite{GHK2}] \label{lemqgood}
 Fix $E_{0}>0$ and consider an energy $E \in [0,E_{0}]$.  Suppose the box $\Lambda=\Lambda_{L}$ (with $L$ large)
is $(X,E,m)$-good for some $X \in  \cQ_{\Lambda_{L}}^{(0\prime)}$.  Then for all $Y\in  [X]_{\Lambda}^{\prime}$  the box $\Lambda$
is   $(Y,E,m)$-jgood, that is,
\begin{align}\label{qweg}
\| R_{Y,\Lambda}(E) \|& \le \e^{L^{1-} +\eta_{L}^{\frac 1 4}}\sim 
 \e^{L^{1-} }
\intertext{and} 
\| \chi_x R_{Y,\Lambda}(E) \chi_y \|& \le \e^{-m |x-y|} +
\eta_{L}^{\frac 1 4}\sim  \e^{-m |x-y|}, \; \;  \text{for $x,y \in \Lambda$ with $ |x-y|\ge \tfrac L {10}$}. \label{qgood}
\end{align}
Moreover, if  $X,Y,X\sqcup Y\in  \cQ_{\Lambda}^{(0\prime)}$ and the box $\Lambda$
is $(X,Y,E,m)$-good, then for all $X_{1}\in  [X]_{\Lambda}^{\prime}$ and $Y_{1}\in  [Y]_{\Lambda}^{\prime}$   we have
$X_{1}\sqcup Y_{1}  \in [X\sqcup Y]_{\Lambda}^{\prime}$, and
 the box $\Lambda$
is  $(X_{1},Y_{1},E,m)$-jgood  as in \eqref{qweg} and \eqref{qgood}.
\end{lemma}s

We also have a lemma \cite[Lemma~3.8]{GHK2} about  the distance to the spectrum inside equivalence classes.
 
\begin{lemma}[\cite{GHK2}] \label{lemqgood7}
  Fix $E_{0}>0$ and consider an energy $E \in [0,E_{0}]$ and a box $\Lambda=\Lambda_{L}$ (with $L$ large).  
 Suppose $\dist (E, \sigma(H_{X,\Lambda}))  \le \tau_{L} $  for some $X \in  \cQ_{\Lambda_{L}}^{(0\prime)}$, where $  \sqrt{\eta_{L}}\ll  \tau_{L} < \frac 1 2$. Then
 \beq
\dist (E, \sigma(H_{Y,\Lambda}))  \le \e^{\eta_{L}^{\frac 1 4}} \tau_{L} \quad \text{for all}\quad Y\in  [X]_{\Lambda}^{\prime}.
\eeq
\end{lemma}

 In view of \eqref{totalN}-\eqref{tinyN} we  have
\beq 
\cQ_{\Lambda}^{(0)}\slash \overset{\Lambda}{\sim}\  = \{[J]_{\Lambda};\, J\in  \cJ_{\Lambda}\}, \quad \text{where} \quad \cJ_{\Lambda}:=\{ J \subset     \J_{\Lambda}; \,
\eqref{totalN1} \mbox{ holds}\}, \label{eqclasses}
\eeq
and we can write $\cQ_{\Lambda}^{(0)}$ and  $\Omega_{\Lambda}^{(0)}$ as 
\beq
\cQ_{\Lambda}^{(0)}=\bigsqcup_{J\in  \cJ_{\Lambda}}[J]_{\Lambda} \quad\text{and} \quad \Omega_{\Lambda}^{(0)}=\bigsqcup_{J\in  \cJ_{\Lambda}}\{ \Y \in[J]_{\Lambda}\}.
\eeq

 We now introduce the basic Poisson configurations and basic events with which we will construct all the relevant probabilistic events. These basic events have to take into account in their very structure the presence of free sites and the finite volume reduction. The following definitions are borrowed from \cite{GHK2}.

\begin{definition} Given $\Lambda=\Lambda_{L}(x)$,  a $\Lambda$-bconfset (basic configuration set) is a subset of $ \cQ_{\Lambda}^{(0)}$ of the form
\beq \label{cylinderset}
C_{\Lambda,B,S}:=\bigsqcup_{\eps_{S}\in \{0,1\}^{S}}[B \cup \cX(S,\eps_{S})]_{\Lambda} = \bigsqcup_{S^{\prime}\subset S} [B \cup S^{\prime}]_{\Lambda},
\eeq
 where we always implicitly assume  $B \sqcup S\in  \cJ_{\Lambda}$. 
$C_{\Lambda,B,S}$ is a $\Lambda$-dense bconfset if
  $S$ satisfies the density condition (cf. \eqref{lambhat})
  \begin{equation}  \label{densitycond}
\# (S\cap {\widehat{\Lambda}_{L^{1-}}} )\ge L^{d-} \quad \text{for all
boxes}\quad   \Lambda_{L^{1-}}\subset \Lambda_{L}.
\end{equation}
We also set
\beq  \label{S=empty}
C_{\Lambda,B}:=C_{\Lambda,B,\emptyset}=[B]_{\Lambda} .
\eeq
\end{definition}

\begin{definition} Given $\Lambda=\Lambda_{L}(x)$,  a $\Lambda$-bevent (basic event) is  a  subset of  $ \Omega_{\Lambda}^{(0)}$ of the form
\beq \label{bevent}
\cC_{\Lambda,B,B^{\pr},S}:= \{\Y \in [B \sqcup B^{\pr}\sqcup S]_{\Lambda}\}\cap
\{\X \in C_{\Lambda,B,S}\}  \cap \{\X^{\pr} \in C_{\Lambda,B^{\pr},S}\},
\eeq
 where we always implicitly assume  $B \sqcup B^{\pr}\sqcup S\in  \cJ_{\Lambda}$. In other words, the $\Lambda$-bevent
$\cC_{\Lambda,B,B^{\pr},S}$  consists of all  $\omega \in \Omega_{\Lambda}^{(0)}$  satisfying
 \begin{align}
\begin{array}{ccl}
N_{\X(\omega)}(\Lambda_{\eta_{L}}(j ))=1 & \text{if} & j\in B ,\\
N_{\X^{\pr}(\omega)}(\Lambda_{\eta_{L}}(j ))=1 & \text{if} & j\in B^{\pr} ,\\
N_{\Y(\omega)}(\Lambda_{\eta_{L}}(j ))=1 & \text{if} & j\in S, \\
N_{\Y(\omega)}(\Lambda_{\eta_{L}}(j ))=0 & \text{if} & j\in  \J_{\Lambda}\backslash (B\sqcup B^{\pr}\sqcup S).
\end{array}
\end{align}
$\cC_{\Lambda,B,B^{\pr},S}$ is a $\Lambda$-dense bevent if
  $S$ satisfies the density condition \eqref{densitycond}. In addition, we set
\beq  \label{S=empty2}
\cC_{\Lambda,B,B^{\pr}}:=\cC_{\Lambda,B,B^{\pr},\emptyset}=\{\X \in C_{\Lambda,B}\}  \cap \{\X^{\pr} \in C_{\Lambda,B^{\pr}}\}.
\eeq
\end{definition}

The number of possible bconfsets and bevents in a given box is always finite. 
We always have 
\beq
\cC_{\Lambda,B,B^{\pr}, S}\subset \{\X \in  C_{\Lambda,B,S}\}\cap \Omega_{\Lambda}^{(0)},
\eeq
\beq
\cC_{\Lambda,B,B^{\pr},S}\subset
\cC_{\Lambda,\emptyset,\emptyset,B\sqcup B^{\pr}\sqcup S}= \{ \Y \in [B\sqcup B^{\pr}\sqcup S]_{\Lambda}\}.
\eeq
 Note also that it follows from \eqref{defOmega0}, \eqref{eqclasses} and \eqref{S=empty2} that
\beq  \label{unionOmega0}
\Omega_{\Lambda}^{(0)}= \bigsqcup_{\{(B,B^{\pr}); \, B\sqcup B^{\pr}\in\cJ_{\Lambda}\} } \cC_{\Lambda,B,B^{\pr}}
\eeq
Moreover,  for each $S_{1}\subset S$ we  have 
\begin{align} \label{cylinderexp}
C_{\Lambda,B,S}&= \bigsqcup_{S_{2}\subset  S_{1}} C_{\Lambda,B \sqcup S_{2},S\setminus S_{1}},\\
\cC_{\Lambda,B,B^{\pr},S}&= \bigsqcup_{S_{2}\subset  S_{1}} \cC_{\Lambda,B \sqcup S_{2}, B^{\pr}\sqcup (S_{1}\setminus S_{2}) ,S\setminus S_{1}}. \label{cylinderexp2}
\end{align}

 Lemma~\ref{lemqgood} leads to  the following definition.

\begin{definition} \label{defadapted}
 Consider an energy $E \in \R$, $m>0$, and a box 
  $\Lambda=\Lambda_{L}(x)$.  
The $\Lambda$-bevent $\cC_{\Lambda,B,B^{\pr},S}$ and the $\Lambda$-bconfset $C_{\Lambda,B,S}$  are  $(\Lambda,E,m)$-good if the
box $\Lambda$ is $(B,S,E,m)$-good.   \emph{(Note that $\Lambda$ is then  $(\omega,E,m)$-jgood for every $\omega \in \cC_{\Lambda,B,B^{\pr},S}$.)}  Those
$(\Lambda,E,m)$-good bevents and bconfsets that are also $\Lambda$-dense will be called  $(\Lambda,E,m)$-adapted.     
\end{definition}

\begin{definition}  Consider an energy $E \in \R$, a rate of decay  $m>0$, and a box   $\Lambda$.  We call $\Omega_{\Lambda}$ a  $(\Lambda,E,m)$-localized event if
there exist  disjoint  $(\Lambda,E,m)$-adapted  bevents
	$\{\cC_{\Lambda,B_{i},B_{i}^{\pr},S_{i}}\}_{i=1,2,\ldots,I}$  such that 
\beq \label{adapted3}
\Omega_{\Lambda}= \bigsqcup_{i=1}^{I} \cC_{\Lambda,B_{i},B_{i}^{\pr},S_{i}} .
\eeq
\end{definition}

If  $\Omega_{\Lambda}$ is a  $(\Lambda,E,m)$-localized event, note  that $\Omega_{\Lambda} \subset \Omega_{\Lambda}^{(0)}$ by its definition, and hence, recalling \eqref{cylinderexp2}  and   \eqref{S=empty2} , we can rewrite $\Omega_{\Lambda} $ in the form
 \beq \label{goodcylinders}
\Omega_{\Lambda}= \bigsqcup_{j=1}^{J} \cC_{\Lambda,A_{j},A_{j}^{\pr}},
\eeq
where the $\{\cC_{\Lambda,A_{j},A_{j}^{\pr}}\}_{j=1,2,\ldots,J}$ are disjoint $(\Lambda,E,m)$-good  bevents.

We will need $(\Lambda,E,m)$-localized events of scale appropriate probability.

\begin{definition}  Fix $p>0$. Given an energy $E \in \R$ and  a rate of decay  $m>0$,  a  scale $L$ 
 is $(E,m)$-localizing if for some box  $\Lambda=\Lambda_{L}$ (and hence for all) 
 we have a $(\Lambda,E,m)$-localized event $\Omega_{\Lambda}$ such that 
\beq \label{PEL0}
\P\{\Omega_{\Lambda}\} > 1- L^{-p}.
\eeq
\end{definition}

 \section{``A priori'' finite volume estimates}  \label{sectinit}
 
Given an energy $E$, to start the multiscale analysis we will need, as in \cite{B,BK},  an {\it a priori}   estimate on the probability that a box $\Lambda_{L}$ is good with an adequate supply of free sites,
for some sufficiently large scale $L$. The multiscale analysis will then show that such a probabilistic estimate also holds at all large scales.

\begin{proposition}\label{proppoisson}  Let $H_{\X}$ be an attractive  Poisson Hamiltonian on  $\L^{2}(\R^{d})$ with density $\vrho >0$, and  fix $p>0$. Then  there exists
 a scale  $\overline{L}_{0}=\overline{L}_{0}(d,u,\vrho,p) < \infty$,  such that for all scales $L\ge\overline{ L}_{0}$,
setting
\begin{gather}  \label{delta0}
 \delta_L:=( \vrho^{-1}(p+d+1) \log L)^{\frac1d}, \quad E_{L}:=  - 2{u_+} \vrho\log L,
\\
\intertext{and}
 m_{L,E}:= \tfrac 1 2 \sqrt{ \tfrac 1 2 E_{L}-E} \le  m_{L}:= \tfrac 1 2 \sqrt{-  \tfrac 1 2 E_{L}} \quad \text{for all $E \in ]-\infty,E_{L}]$}, \label{mL}
\end{gather}
the  scale $L$  is $(E,m_{L,E})$-localizing for all energies $E \in ]-\infty,E_{L}]$.
\end{proposition}

\begin{proof}  Let $\Lambda=\Lambda_{L}(x)$, and let $\delta_{L}$ and $ E_{L}$  be as in \eqref{delta0}.   If $Y \in \cQ^{(0)}_{\Lambda}$, it follows from  \eqref{lowerboundinbox} and the  Combes-Thomas estimate (e.g.,  \cite[Eq. (19)]{GKkernel}) that 
 for all $E \in ]-\infty, E_{L}]$ and all $t_{Y} \in [0,1]^{Y}$ we have, with $m_{L}(E)$
 as in \eqref{mL}, that
\begin{align}\label{wegL0}
\| R_{(Y,t_{Y}),\Lambda}(E) \|& \le (2 m_{L,E})^{-2}\\
\intertext{and}   \label{goodL01}
\| \chi_y  R_{(Y,t_{Y}),\Lambda}(E)  \chi_{y^{\prime}} \|& \le 
\tfrac 12   m_{L,E}^{-2}\, \e^{-2 m_{L,E}|y-y^{\prime}|}\;  \text{for $y,y^{\prime} \in \Lambda$ with $\abs{y-y^{\pr}} \ge 4 \sqrt{d}$}.
 \end{align}
 
We now require
$L> \delta_{L}+ \delta_{+}$,  and set
\begin{align}\label{setJ}
   J&:=\{j \in x + \delta_L \Z^d \cap\Lambda; \,  \Lambda_{\delta_L}(j)\subset \widehat{\Lambda}\},\\ \label{NSjpp}
\widehat{\cJ}_{\Lambda}&:=\{S \in{\cJ}_{\Lambda}; \,  N_{S}(\Lambda_{\delta_L}(j))\ge 1 \;\text{for all} \; j\in J\}.
\end{align}
If $S \in \widehat{\cJ}_{\Lambda} $, the density condition \eqref{densitycond} for $S$ in $\Lambda$  follows from \eqref{NSjpp}, and it follows from \eqref{wegL0} and  \eqref{goodL01}  that  $ \cC_{\Lambda,\emptyset,\emptyset,S}$ is a $(\Lambda,E,m_{L,E})$-adapted bevent  for all  $E \in ]-\infty, E_{L}]$  if $L\ge\overline{ L}_{1}(d,u,\vrho,p)$. 
  We conclude that
\beq 
\Omega_{\Lambda}=\bigsqcup_{S\in \widehat{\cJ}_{\Lambda}} \cC_{\Lambda,\emptyset,\emptyset,S}=\bigsqcup_{S\in \widehat{\cJ}_{\Lambda}} \{\Y\in [S]_{\Lambda} \}
\eeq
is a $(\Lambda,E,m_{L,E})$-localizing event for all  $E \in ]-\infty, E_{L}]$. 

To establish \eqref{PEL0}, let $\delta_{L}^{\pr}:=\frac 1 {2^{d}}\delta_{L}$ and consider the event
\beq
\Omega_{\Lambda}^{(\ddagger)}:=\{ N_{\Y}(\Lambda_{\delta_{L}^{\pr}}(j))\ge 1 \quad \text{for all} \quad  j \in J \}.
\eeq
  We have
 \begin{equation}\label{PElarge}
\P \{\Omega_{\Lambda}^{(\ddagger)} \}\ge 1 - \left(\tfrac L{\delta_L}\right)^d \e^{-2\vrho (\delta_L^{\pr})^d}
= 1 - \frac{\vrho}{(p+d+1) L^{p+1} \log L} \ge 1 - \frac1{ L^{p+1}},
\end{equation}
if $L\ge \e^{\frac {\vrho}{p+d+1}}$. Since
$\Omega_{\Lambda} \supset \Omega_{\Lambda}^{(\ddagger)} \cap \Omega_{\Lambda}^{(0)}$, \eqref{PEL0} follows from \eqref{PElarge} and \eqref{Omega0} for ${L}\ge\overline{L}_{0}(d,u,\vrho,p)$.
\end{proof}

\section{The multiscale analysis and the proof of localization}
\label{sectMSA}

The Bourgain-Kenig multiscale analysis, namely 
\cite[Proposition~A$^{\!\prime}$]{BK}, was adapted to Poisson Hamiltonians in \cite[Proposition~5.1]{GHK2}.  To apply the latter  to attractive Poisson Hamiltonians we must show  that the requirements of this multiscale analysis are satisfied.  More precisely,
 we must show that attractive Poisson Hamiltonians  satisfy appropriate versions of Properties SLI
 (Simon-Lieb inequality), EDI (eigenfunction decay inequality),  IAD
(independence at a distance), NE (average number of eigenvalues),  and GEE (generalized eigenfunction expansion); see   \cite{GK1}. The Wegner estimate is proved by the multiscale analysis; it is not an ``a priori requirement''.

 Since events based on disjoint boxes are independent, we have Property IAD.  Property GEE is  satisfied in view of \eqref{HK}, and we also have  \eqref{HK2}, which is needed in the multiscale analysis.

But Properties SLI, EDI and NE require some care and modification.  In a box $\Lambda_{L}$ we always work with $\Lambda_{L}$-acceptable configurations $X$, whence the potential $V_{X,\Lambda_{L}}$ satisfies the lower bound \eqref{controlV2}.  Inside the box  $ \Lambda_{L}$, Properties   SLI and EDI (see \cite[Theorem~A.1]{GKduke}, \cite[Section~2]{BK}) are governed by the same constant $\gamma_{E,L}$ given in  \cite[Eq.~(A.2)]{GKduke}, and hence for  $\Lambda_{L}$-acceptable configurations we have
\beq \label{SLIEDI}
\gamma_{]-\infty,0],L}:= \sup_{E \in ]-\infty,0],L} \gamma_{E,L}  \le
C_{d}\sqrt{ u_{+} \vrho \log L}.
 \eeq
For Property NE, it follows  by the argument in  \cite[Eqs. (A.6)-(A.7)]{GKduke}
that for all $\Lambda_{L}$-acceptable configurations $X$ and energies $E \in  ]-\infty,0]$
we have
\begin{align}\label{NE}
\mathrm{tr}\,\left\{ \chi_{(-\infty,E)} (H_{X,\Lambda_{L}})\right\} &\le 
\mathrm{tr}\,\left\{
 \chi_{\left(-\infty,{E + u_{+} \vrho \log L}\right)}
 (-\Delta_{\Lambda_{L}})   \right\} \\
&\le C_d 
\left( u_{+} \vrho \log L\right)^{\frac d2} L^d \nonumber .
\end{align}
The extra factors of $\sqrt{\log L}$ in \eqref{SLIEDI} and \eqref{NE} are acceptable in the multiscale analysis.

The Wegner estimate is proved at each scale using  \cite[Lemma~5.1$^{\pr}$]{BK}. The sign of the single-site potential does not matter in this argument, as long as the single-site potential  has a definite sign, positive or negative, to ensure the monotonicity of the eigenvalues in the free sites couplings.

Thus the following proposition  follows from  Proposition~\ref{proppoisson} and \cite[Proposition~5.1]{GHK2}.

\begin{proposition}\label{propAB}  Let $H_{\X}$ be an attractive Poisson Hamiltonian on  $\L^{2}(\R^{d})$ with density $\vrho >0$ and $p=\frac 3 8 d -$.  Then
there exist an energy $E_{0}=E_0(\vrho)<0$ and  
 a scale $L_{0}=L_{0}(\vrho)$, such that setting
$m_{E} := \tfrac 1 4 \sqrt{ \tfrac 1 2 E_{0}-E} \le  m_{E_{0}}:= \tfrac 14 \sqrt{-  \tfrac 1 2 E_{0}}$, the scale
  $L$ is $(E,m_{E})$-localizing for all $L \ge L_{0}$ and $E \in ]-\infty,E_{0}]$.
\end{proposition}

Theorem~\ref{thmpoisson} now follows from Proposition~\ref{propAB} as in \cite[Proposition~6.1]{GHK2}.

\end{document}